\documentclass[pre,aps,twocolumn,superscriptaddress]{revtex4-1}

\usepackage{graphicx}
\usepackage{amssymb,amsfonts,amsmath}
\usepackage{color}
\usepackage{ulem}
\usepackage[hidelinks]{hyperref}
\usepackage{bbm}
\usepackage{bbold}

\usepackage{tikz}
\usetikzlibrary{shapes}
\usetikzlibrary{patterns}
\usetikzlibrary{angles, quotes}

\newcommand{\rv}{{\mathbf r}}

\newcommand{\Tr}{{\rm Tr}\,}

\newcommand{\Jv}{{\bf J}}

\newcommand{\fv}{{\bf f}}

\newcommand{\vel}{{\bf v}}

\newcommand{\msphantom}[1]{$\ldots$}

\newcommand{\exc}{{\rm exc}}

\newcommand{\mydelete}[1]{{}}

\newcommand{\rmint}{{\rm int}}
\newcommand{\rmad}{{\rm ad}}
\newcommand{\rmsup}{{\rm sup}}
\newcommand{\rmexc}{{\rm exc}}
\newcommand{\rmext}{{\rm ext}}

\newcommand{\rmid}{{\rm id}}

\newcommand{\fext}{\fv^\rmext}
\newcommand{\Vext}{V^\rmext}
\newcommand{\Vextdot}{\dot V^\rmext}
\newcommand{\pairpotential}{\phi}

\begin{document}

\title{Entropy power functional theory for Brownian many-body dynamics}

\author{Matthias Schmidt}
\affiliation{Theoretische Physik II, Physikalisches Institut, 
  Universit{\"a}t Bayreuth, D-95447 Bayreuth, Germany}
\email{Matthias.Schmidt@uni-bayreuth.de}

\begin{abstract}
We present a formally exact variational scheme for the overdamped
Brownian dynamics of pairwise interacting many-body systems in general
spatiotemporal nonequilibrium. A joint free power minimization
principle determines instantaneously the one-body current and the
global interparticle distance flux. The intrinsic free power
functional splits into entropic and energetic rates, where the latter
are treated explicitly.  The adiabatic contribution to the entropy
rate is the time derivative of the equilibrium entropy metadensity
functional. Genuine nonequilibrium effects originate from a universal
entropy superpower functional. Two continuity equations close the
dynamical description.
\end{abstract}

\date{21 July 2025}

\maketitle

Power functional theory is a formally exact variational framework for
the nonequilibrium statistical mechanics of many-body systems
\cite{schmidt2022rmp}. The theory provides functional closure on the
one-body level of dynamical correlation functions. Higher-body
correlation functions are systematically accessible via dynamical test
particle \cite{schmidt2022rmp, brader2015dtpl,
  treffenstaedt2021dtpl,treffenstaedt2022dtpl} and nonequilibrium
Ornstein-Zernike routes \cite{schmidt2022rmp, brader2013noz,
  brader2014noz}. Concrete formulations were given for overdamped
Brownian dynamics~\cite{schmidt2013pft}, classical molecular dynamics
\cite{schmidt2018md}, and quantum many-body
dynamics~\cite{schmidt2015qpft}. The nonequilibrium minimization
principle implies computational and conceptual efficiency on the
dynamical one-body level.
In equilbrium the approach reduces to density functional theory
\cite{hansen2013, evans1979, evans1992, evans2016}, where the
`adiabatic' interparticle force field is accessible via the gradient
of the equilibrium one-body direct correlation functional. Systems
that are spatially inhomogeneous, from molecular to macroscopic
scales, can be addressed and the power functional generalizes this
property to the temporal domain with general time dependence.

The behaviour of the relevant `superadiabatic' force functionals was
studied in a variety of physical contexts and for different
systems. For overdamped Brownian dynamics, systematic work was
addressed at one-dimensional systems \cite{fortini2014prl,
  bernreuther2016gcm} and later shown to generalize to higher
dimensions via velocity gradient approximations
\cite{delasheras2018velocityGradient}. Superadiabatic forces were
classified into nonequilibrium structural and flow parts
\cite{stuhlmueller2018prl, delasheras2020fourForces}. Memory-induced
motion reversal \cite{treffenstaedt2020shear} and the response of
fluids \cite{jahreis2019shear} and gels \cite{treffenstaedt2020shear}
to shear were studied.  The time-dependent two-body correlation
structure was analyzed on the basis of the van Hove function
\cite{brader2015dtpl, treffenstaedt2021dtpl,treffenstaedt2022dtpl}. In
counterdriven binary mixtures, demixing \cite{geigenfeind2020laning}
and nonequilibrium drag force scaling \cite{koeglmayr2024scaling} were
investigated.  The physics of active Brownian particles was addressed,
including nonequilibrium phase coexistence \cite{krinninger2016prl,
  krinninger2019jcp, hermann2024activeFreezing}, the mechanical
balance of the nonequilibrium force fields \cite{hermann2019prl,
  hermann2021molPhys}, and the unambiguous positive interfacial
tension for motility-induced phase separation \cite{hermann2019prl}.
Acceleration viscous effects were identified in quantum
\cite{bruetting2019viscosity} and classical \cite{renner2022prl}
Hamiltonian dynamics.

The dynamical power functional map generalizes the Mermin-Evans
equilibrium density functional theorem~\cite{mermin1965, evans1979},
which establishes that in principle one is able to determine uniquely,
given the density profile and thermodynamic parameters, the
corresponding external one-body potential. The practical validity of
the nonequilibrium generalization \cite{schmidt2022rmp} was
demonstrated via simulation-based custom flow methods both for
Brownian \cite{delasheras2019customFlow} and molecular
\cite{renner2021customFlowMD} dynamics. Further standalone evidence
was provided from a data-driven standpoint via the successful machine
learning of superadiabatic force functionals in steady states
\cite{delasheras2023perspective, zimmermann2024ml} and under
time-dependent active swimming \cite{fleischmann2026}. Functional
machine learning is a powerful new approach that has sparked much
progress in the equilibrium physics of soft
matter~\cite{sammueller2023neural,dijkman2024ml,sammueller2024attraction,
  bui2024neuralrpm, bui2025dielectrocapillarity, bui2026abinitio,
  sammueller2024hyperDFT, kampa2024meta, kampa2026pairmatching,
  kampa2026spherical}.  Power functional theory allows one to overcome
systematically the limitations of dynamical density functional theory
\cite{evans1979, marconi1999, marconi2007, schilling2022,
  delasheras2023perspective}. Close interconnections exist with
Noether's theorem~\cite{noether1918, hermann2021noether,
  mueller2024gauge, mueller2024whygauge} and with the very recent
dynamical statistical mechanical gauge
invariance~\cite{mueller2024dynamic}.

Here we develop the entropic version of power functional theory, in
generalization of the recent entropic density functional theory for
the physics of spatially inhomogeneous systems in thermal
equilibrium~\cite{schmidt2026entropyShort,
  schmidt2026entropyLong}. The present nonequilibrium approach is
based on an extended exact joint free power minimization principle,
which allows one to split the free power systematically into energetic
and entropic contributions. The variational fields are the position-
and time-dependent one-body density, $\rho(\rv,t)$, and current,
$\Jv(\rv,t)$, which are both standard in Brownian
dynamics~\cite{schmidt2022rmp} and here partnered by the inclusion of
the time-dependent global distance distribution, $G(r,t)$, and the
associated global distance flux, $J_G(r,t)$.  Temporal entropic
changes are uniquely identified on the functional level and these
split into adiabatic and genuine nonequilibrium contributions. The
adiabatic part is the time derivative of the equilibrium excess
entropy metadensity functional $\dot S_\rmexc[\rho,G]$
\cite{schmidt2026entropyShort, schmidt2026entropyLong}. The genuine
nonequilibrium part is the superadiabatic excess entropy power
(`superpower') functional~$\Sigma_t[\rho,G,\Jv,J_G]$.

We consider nonequilibrium statistical mechanics as given on the
many-body level by the overdamped dynamics of the probability
distribution function $\Psi(\rv^N,t)$. Here $\rv^N=\rv_1,\ldots,\rv_N$
is a shorthand for all $N$ particle position coordinates. The time
evolution is determined by the Smoluchowski equation
\cite{zwanzig2001, dhont1996book}, $\dot \Psi(\rv^N,t) = -\sum_i
\nabla_i \cdot \vel_i \Psi(\rv^N,t)$, where the overdot indicates the
partial derivative with respect to time $t$, the sum runs over all
particle indices $i=1,\ldots, N$, and $\nabla_i$ denotes the
derivative with respect to position $\rv_i$ of
particle~$i$~\cite{schmidt2022rmp}.
For given particle configuration~$\rv^N$, the velocity of particle $i$
is given at time~$t$ by the following (many-body level) vector field:
\begin{align}
  \gamma \vel_i(\rv^N,t) = -\nabla_i u(\rv^N,t) +
  \fext(\rv_i,t) -\nabla_i \ln\Psi(\rv^N,t).
  \label{EQveliDefinition}
\end{align}
Here $\gamma$ is the friction constant against the background, the
interparticle interaction potential $u(\rv^N,t)$ is in general
explicitly dependent on time and further specified below, and
$\fext(\rv,t)$ is an external force field, expressed as a function of
a generic position vector $\rv$ and time~$t$.

We take the interparticle interaction potential to consist of only
pairwise contributions, $u(\rv^N,t) = \sum_{i,j\neq
  i}\pairpotential(|\rv_i- \rv_j|,t)/2$, where $\pairpotential(r,t)$
is the pair potential as a function of (scalar) interparticle distance
$r$ and time~$t$.  The double sum over distinct particle pairs is
indicated by $\sum_{i,j\neq i}=\sum_{i=1}^N \sum_{j=1, j\neq i}^N$.
The external force field $\fext(\rv,t)$ consists in general of
separate conservative and nonconservative contributions.  The
conservative part can be expressed as $-\nabla \Vext(\rv,t)$, where
$\Vext(\rv,t)$ is the external potential and $\nabla$ indicates the
derivative with respect to~$\rv$.
At the initial time $t=0$ the system is taken to be in thermal
equilibrium under the influence of the external potential
$V_\rmext(\rv,0)$. Time-dependent averages $A(t)$ of observables $\hat
A(\rv^N)$ are built according to the standard
procedure~\cite{zwanzig2001, dhont1996book}, $A(t)=\langle \hat
A(\rv^N);t\rangle$, where the angle brackets denote the instantaneous
dynamical average $\langle\,\cdot\,;t\rangle = \Tr
\,\cdot\,\Psi(\rv^N,t)$. Here the configurational `trace' denotes the
integral over all position coordinates, $\Tr\,\cdot\,=\int
d\rv_1\ldots d\rv_N$, and the many-body distribution function is
normalized according to $\Tr \Psi(\rv^N,t)=1$.

We use one-body observables to construct a functional description.
The standard density operator $\hat\rho(\rv)$ \cite{evans1979,
  hansen2013, schmidt2022rmp} and the `instantaneous pair distance
histogram' $\hat G(r)$ \cite{kampa2024meta, kampa2026pairmatching,
  kampa2026spherical, schmidt2026entropyShort, schmidt2026entropyLong}
are given respectively by
\begin{align}
  \hat\rho(\rv) &= \sum_i \delta(\rv-\rv_i),
  \label{EQrhoHatDefinition}\\
  \hat G(r) &= \frac{1}{2}  \sum_{i,j\neq i}\delta(r-|\rv_i-\rv_j|),
  \label{EQGhatDefinition}
\end{align}
where $\delta(\,\cdot\,)$ indicates the Dirac distribution.  To
characterize the motion, we use the standard one-body current operator
$\hat\Jv(\rv)$~\cite{schmidt2022rmp} and complement this by a
corresponding `distance current' operator $\hat J_G(r)$. The two
explicit definitions are:
\begin{align}
  \hat \Jv(\rv) &= \sum_i \delta (\rv-\rv_i) \vel_i,
  \label{EQcurrentOperatorDefinition}\\
  \hat J_G(r) &= \sum_{i,j\neq i}
  \delta(r-|\rv_i-\rv_j|) 
  \frac{\rv_i-\rv_j}{|\rv_i-\rv_j|} \cdot (\vel_i - \vel_j),
  \label{EQdistanceCurrentOperatorDefinition}
\end{align}
where $\vel_i(\rv^N,t)$ is the many-body velocity
\eqref{EQveliDefinition}.  Averaging the observables
\eqref{EQrhoHatDefinition}--\eqref{EQdistanceCurrentOperatorDefinition}
yields the dynamical density profile $\rho(\rv,t)=\langle
\hat\rho(\rv);t \rangle$, the mean distance histogram $G(r,t)=\langle
\hat G(r);t \rangle$, the one-body current density $\Jv(\rv,t)=\langle
\hat\Jv(\rv);t \rangle$, and the global distance current distribution
function $J_G(r,t)=\langle \hat J_G(r);t \rangle$.

It is straightforward to show that the standard continuity equation
\eqref{EQcontinuity1} for position holds and that it is partnered with
a continuity equation \eqref{EQcontinuity2} for pair distance:
\begin{align}
  \dot\rho(\rv,t) &= -\nabla \cdot \Jv(\rv,t),
  \label{EQcontinuity1}\\
  \dot G(r,t) &= -\partial_r J_G(r,t),
  \label{EQcontinuity2}
\end{align}
where we recall that the overdot indicates $\partial/\partial t$, the
position derivative is $\nabla=\partial/\partial \rv$, and we denote
the partial derivative with respect to distance by $\partial_r =
\partial / \partial r$.

In generalization of the standard power functional
dependence~\cite{schmidt2022rmp} on $\rho(\rv,t)$ and $\Jv(\rv,t)$,
here we consider the free power functional $R_t[\rho, G, \Jv, J_G]$ to
acquire further functional dependence on the distance distribution
$G(r,t)$ and on the distance current $J_G(r,t)$.
The free power consists of a sum of an external part,
$X_t^\rmext[\rho,\Jv]$, and an intrinsic
contribution~\cite{schmidt2022rmp}. The present formulation allows us
to split the intrinsic contribution, i.e.\ the difference $R_t[\rho,
  G, \Jv, J_G] - X_t^\rmext[\rho,\Jv]$, furter into an explicit
energetic part, $X_t^\rmint[G,J_G]$, and an intrinsic remainder,
$W_t^\rmint[\rho,G,\Jv,J_G]$. Thus the power functional consists of
the following three terms:
\begin{align}
  R_t[\rho,G,\Jv,J_G] &= 
  W_t^\rmint[\rho,G,\Jv,J_G]
  \notag\\&\quad
  -X_t^\rmint[G,J_G] - X_t^\rmext[\rho,\Jv].
  \label{EQRtDefinition}
\end{align}
The two latter contributions, $X_t^\rmint[G,J_G]$ and
$X_t^\rmext[\rho,\Jv]$, are purely energetic and given by the
following instantaneous (position and distance) integrals:
\begin{align}
  X_t^\rmext[\rho,\Jv] &= 
  \int d\rv
  \big[\Jv(\rv,t) \cdot \fext(\rv,t) 
    - \rho(\rv,t)\Vextdot(\rv,t)\big],
  \label{EQXextDefinition} \\
  X_t^\rmint[G,J_G] &= 
  -\int dr \big[J_G(r,t) \pairpotential'(r,t) 
    + G(r,t) \dot\pairpotential(r,t)\big].
  \label{EQXintDefinition}
\end{align}
The integrands in Eqs.~\eqref{EQXextDefinition} and
\eqref{EQXintDefinition} constitute sums of power densities that arise
from `mechanical' particle motion (first term in each integral) and
static `charging' due to the temporal change of the respective
potential energy function (second term in each integral).

The extended power functional variational principle implies the
simultaneous instantaneous minimization with respect to both currents
$\Jv(\rv,t)$ and $J_G(r,t)$.
As a consequence, two Euler-Lagrange equations hold:
\begin{align}
  \frac{\delta R_t[\rho,G,\Jv,J_G]}{\delta\Jv(\rv,t)} 
  &= 0 \qquad \text{(min)},
  \label{EQpowerEL1}  \\
  \frac{\delta R_t[\rho,G,\Jv,J_G]}{\delta J_G(r,t)} &= 0 \qquad
  \text{(min)}.
  \label{EQpowerEL2}
\end{align}
In the following we lay out that the Euler-Lagrange equations
\eqref{EQpowerEL1} and \eqref{EQpowerEL2} constitute nonequilibrium
force balance relationships, which together with the continuity
equations \eqref{EQcontinuity1} and \eqref{EQcontinuity2} determine
formally the full dynamics of the system.

As a demonstration of the general concept to generate force fields
from power functional differentiation~\cite{schmidt2022rmp}, we first
address the two potential force contributions. Due to the simple
linear dependence of the energy rate functionals
\eqref{EQXextDefinition} and \eqref{EQXintDefinition} on the
respective currents $\Jv(\rv,t)$ and $J_G(r,t)$, it is straightforward
to obtain the functional derivatives, which are relevant for the
Euler-Lagrange equations \eqref{EQpowerEL1} and
\eqref{EQpowerEL2}. Functionally differentiation yields the following
simple forms:
\begin{align}
  \frac{\delta X_t^\rmext[\rho,\Jv]}{\delta \Jv(\rv,t)}
  &=  \fv^\rmext(\rv,t),
  \label{EQexternalForceFromDifferentiation}  \\
  \frac{\delta X_t^\rmint[G,J_G]}{\delta J_G(r,t)}
  &= -\phi'(r,t).
  \label{EQinterparticleForceFromDifferentiation}
\end{align}
where we recall $-\phi'(r,t)$ as the pair-distance-resolved
interparticle force.

All further force contributions are generated by the intrinsic free
power functional $W_t^\rmint[\rho,G,\Jv,W]$; we recall the functional
splitting \eqref{EQRtDefinition} and further decompose:
\begin{align}
  W_t^\rmint[\rho,G,\Jv,J_G] &= P_t^\rmid[\rho,G,\Jv,J_G] 
  + \dot F_t^\rmid[\rho] 
  \notag\\&\quad
- T \big(\dot S_t^\rmexc[\rho,G]
  + \Sigma_t^\exc[\rho,G,\Jv,J_G] \big),
  \label{EQWtDefinition}
\end{align}
The ideal dissipation functional, $P_t^\rmid[\rho,G,\Jv,J_G]$, and the
time derivative of the intrinsic Helmholtz ideal free energy
functional, $\dot F_t^\rmid[\rho]$, are both local and instantaneous
functionals that are given in closed form below.
All `adiabatic' correlation effects are contained in the time
derivative of the equilibrium excess entropy metadensity functional,
$\dot S_t^\rmexc[\rho,G]$, which is in general nonlocal
\cite{schmidt2026entropyShort, schmidt2026entropyLong} and
instantaneous in time.
The entropic superpower functional, $\Sigma_t^\rmexc[\rho,G,\Jv,J_G]$,
is the genuine nonequilibrium contribution to the free power and this
depends in general causally in time and nonlocally in position on its
functional arguments.  Crucially, both $\dot S_t^\rmexc[\rho,G]$ and
$\Sigma_t^\rmexc[\rho,G,\Jv,J_G]$ are independent of the pair
potential $\pairpotential(r,t)$, which rather features solely in the
intrinsic interparticle power, $X_t^\rmint[G,J_G]$, see its definition
\eqref{EQXintDefinition} and the resulting interparticle force
\eqref{EQinterparticleForceFromDifferentiation}.

To provide context, we recall the standard form~\cite{schmidt2022rmp}
of the ideal dissipation functional as $P_t^\rmid[\rho,\Jv]=\gamma\int
d\rv [\Jv(\rv,t)]^2/[2\rho(\rv,t)]$. The ideal friction force field
follows as the negative functional derivative with respect to the
instantaneous one-body current, $-\delta
P_t^\rmid[\rho,\Jv]/\delta\Jv(\rv,t)=-\gamma \vel(\rv,t)$.  Here the
velocity field is the ratio $\vel(\rv,t)=\Jv(\rv,t)/\rho(\rv,t)$ and
changing the functional variable from $\Jv(\rv,t)$ to $\vel(\rv,t)$
yields the equivalent functional form $P_t^\rmid[\rho,
  \vel]=\gamma\int d\rv\rho(\rv,t) [\vel(\rv,t)]^2/2$.

Resolving the pair distance dynamics on the functional level leads to
additional functional dependence of $P_t^\rmid[\rho,G,\Jv,J_G]$ on
$G(r,t)$ and on $J_G(r,t)$. It is useful to express the thus extended
ideal dissipation functional as the sum of two separate contributions:
\begin{align}
  P_t^\rmid[\rho,G,\Jv,J_G] &= P_t^{\rmid,1}[G,J_G]
  + P_t^{\rmid,2}[\rho,G,\Jv,J_G].
  \label{EQPidDefinition}
\end{align}
The first term, $P_t^{\rmid,1}[G,J_G]$, is a pure distance
contribution with standard quadratic current dependence:
\begin{align}
  P_t^{\rmid,1}[G,J_G] &= 
  \frac{\gamma}{2}\int dr \frac{[J_G(r,t)]^2}{G(r,t)}
  \label{EQPid1current}\\
  &= \frac{\gamma}{2} \int dr G(r,t)[v_G(r,t)]^2,
  \label{EQPid1velocity}
\end{align}
where the alternative form \eqref{EQPid1velocity} follows from using
the distance velocity $v_G(r,t)$, which is defined as the ratio:
\begin{align}
  v_G(r,t) &= \frac{J_G(r,t)}{G(r,t)}.
  \label{EQvGDefinition}
\end{align}
The ideal motion of position and of distance is coupled in the sum
\eqref{EQPidDefinition} via the second term, which is given, in again
two alternative forms, by:
\begin{align}
  P_t^{\rmid,2}[\rho,G,\Jv,J_G] 
  &= \gamma \int dr J_G(r,t) v_G^\rmid(r,t;[\rho,G,\Jv]),
  \label{EQPid2distance}\\
  &= \gamma \int d\rv \Jv(\rv,t)\cdot 
  \vel_\rho^\rmid(\rv,t;[\rho,G,J_G]).
  \label{EQPid2position}
\end{align}
The first form \eqref{EQPid2distance} contains the ideal distance
velocity functional $v_G^\rmid(r,t;[\rho,G,\Jv])$ and corresponding
ideal distance flux functional $J_G^\rmid(r,t;[\rho,\Jv])$, which are
given via:
\begin{align}
  & v_G^\rmid(r,t;[\rho,G,\Jv]) = 
  \frac{J_G^\rmid(r,t;[\rho,\Jv])}{G(r,t)},
  \label{EQvidGDefinition}\\
  &J_G^\rmid(r,t;[\rho,\Jv]) =
  \notag\\&\quad
  \int d\rv \Jv(\rv,t) \cdot  
  \int d\rv' \rho(\rv') \delta(r-|\rv-\rv'|) \frac{\rv-\rv'}{|\rv-\rv'|}.
  \label{EQJGDefinition} 
\end{align}
Inserting the definitions~\eqref{EQvidGDefinition} and
\eqref{EQJGDefinition} into the ideal dissipation functional
\eqref{EQPid2distance} and rearranging leads to the alternative form
\eqref{EQPid2position} upon defining the ideal velocity field
functional as:
\begin{align}
  &\vel_\rho^\rmid(\rv,t;[\rho,G,J_G]) =
  \notag\\&\quad
  \int dr \frac{J_G(r,t)}{G(r,t)} 
  \int d\rv' \rho(\rv') \delta(r-|\rv-\rv'|) \frac{\rv-\rv'}{|\rv-\rv'|}.
  \label{EQvelIdealDefinition}
\end{align}

Functional differentiation of the ideal dissipation functional
\eqref{EQPidDefinition} then yields the corresponding (position and
distance) friction force fields according to:
\begin{align}
  -\frac{\delta P_t^\rmid[\rho,G,\Jv,J_G]}{\delta \Jv(\rv,t)} &= 
  -\gamma \vel_\rho^\rmid(\rv,t;[\rho,G,J_G]),
  \label{EQPtidByPosition}\\
  -\frac{\delta P_t^\rmid[\rho,G,\Jv,J_G]}{\delta J_G(r,t)} &=
  -\gamma v_G(r,t),
  \label{EQPtidByDistance}
\end{align}
where the result \eqref{EQPtidByPosition} for position motion includes
the velocity functional~\eqref{EQvelIdealDefinition} and the result
\eqref{EQPtidByDistance} for distance motion contains $v_G(r,t)$ as is
given by the ratio~\eqref{EQvGDefinition}.

We turn to the adiabatic contributions to the free power functional
\eqref{EQWtDefinition}, $\dot F_t^\rmid[\rho]$ and $-T\dot
S_t^\rmexc[\rho,G]$.  The time derivative of the ideal gas free energy
functional $\dot F_t^\rmid[\rho]$ has the standard
form~\cite{schmidt2022rmp},
\begin{align}
  \dot F_t^\rmid[\rho] &= 
  k_BT \int d\rv \Jv(\rv,t) \cdot \nabla \ln\rho(\rv,t),
  \label{EQFidDefintion}
\end{align}
which generates the diffusive force field via
\begin{align}
  -\frac{\delta \dot
  F_t^\rmid[\rho]}{\delta\Jv(\rv,t)} &= -k_BT
  \nabla\ln\rho(\rv,t).
  \label{EQdiffusiveForceFromDifferentiation}
\end{align}
Similarly, the time derivative of the equilibrium excess entropy
metadensity functional $\dot S_t^\rmexc[\rho,G]$ is given by
\begin{align}
  \frac{\dot S_t^\rmexc[\rho,G]}{k_B} &=
  \int d\rv \Jv(\rv,t)\cdot \nabla c_\rho(\rv,t;[\rho,G])
  \notag\\&\quad
  +\int dr J_G(r,t) \partial_r c_G(r,t;[\rho,G]),
  \label{EQSexcDot}
\end{align}
where $c_\rho(\rv,t;[\rho,G])$ and $c_G(r,t;[\rho,G])$ are adiabatic
entropic correlation functionals
\cite{schmidt2026entropyShort,schmidt2026entropyLong}. These follow
from partial functional differentiation of the equilibrium excess
entropy metadensity functional $S_t^\rmexc[\rho,G]$ via
\cite{schmidt2026entropyShort, schmidt2026entropyLong}:
\begin{align}
  c_\rho(\rv,t;[\rho,G]) &= 
  k_B^{-1}\frac{\delta S_t^\rmexc[\rho,G]}{\delta \rho(\rv,t)},
  \label{EQcRhoDefinition}\\
  c_G(r,t;[\rho,G]) &=
  k_B^{-1}\frac{\delta S_t^\rmexc[\rho,G]}{\delta G(r,t)}.
  \label{EQcGDefinition}
\end{align}
Here the time dependence of the adiabatic direct correlation
functionals~\eqref{EQcRhoDefinition} and \eqref{EQcGDefinition} arises
solely via evaluation of the respective functional derivatives at the
instantaneous functional arguments $\rho(\rv,t)$ and $G(r,t)$, which
is the standard mechanism \cite{schmidt2022rmp}.

Based on the  entropic
direct correlation functionals \eqref{EQcRhoDefinition}
and~\eqref{EQcGDefinition},
it is useful to define adiabatic functionals for the entropic position
force, $\fv_\rho^\rmad(\rv,t;[\rho,G])$, and for the entropic distance
force, $f_G^\rmad(r,t;[\rho,G])$  as:
\begin{align}
  \fv_\rho^\rmad(\rv,t;[\rho,G]) &= k_BT \nabla c_\rho(\rv,t;[\rho,G]),
  \label{EQfadRho}\\
  f_G^\rmad(r,t;[\rho,G]) &=  k_BT \partial_r c_G(r,t;[\rho,G]).
  \label{EQfadG}
\end{align}
Using the expressions \eqref{EQfadRho} and \eqref{EQfadG} for the
entropic force functionals allows one to re-write the adiabatic
entropy time derivative \eqref{EQSexcDot} in the following form:
\begin{align}
  T \dot S_t^\rmexc[\rho,G]  &= 
  \int d\rv \Jv(\rv,t) \fv_\rho^\rmad(\rv,t;[\rho,G])
  \notag\\&\quad
  +\int dr J_G(r,t) f_G^\rmad(r,t;[\rho,G]),
  \label{EQadiabaticEntropyRate}
\end{align}
where the (linear) functional dependence on the two currents
$\Jv(\rv,t)$ and $J_G(r,t)$ remains again suppressed in the notation
on the left hand side.
It is then straightforward to see that the specific forms of the
entropic force fields \eqref{EQfadRho} and \eqref{EQfadG} are
generated from the following current functional derivatives of the
adiabatic entropy rate \eqref{EQadiabaticEntropyRate}:
\begin{align}
  \fv_\rho^\rmad(\rv,t;[\rho,G]) &=
  T\frac{\delta \dot S_t^\rmexc[\rho,G]}{\delta \Jv(\rv,t)},
  \label{EQfadRhoFromDifferentiation}\\
  f_G^\rmad(r,t;[\rho,G]) &=  
  T\frac{\delta \dot S_t^\rmexc[\rho,G]}{\delta J_G(r,t)},
  \label{EQfadGFromDifferentiation}
\end{align}
The positive signs on the right hand sides
\eqref{EQfadRhoFromDifferentiation} and
\eqref{EQfadGFromDifferentiation} are consistent with forces arises
from negative current derivatives of the free power
\cite{schmidt2022rmp}; we recall the further negative sign of the
entropic contribution to the intrinsic free
power~\eqref{EQWtDefinition}.

It remains to address the ultimate term in the intrinsic power
splitting \eqref{EQWtDefinition}, which is the entropic superpower
functional $\Sigma_t^\rmexc[\rho,G,\Jv,J_G]$.  Superadiabatic entropic
force fields are genereated according to the following current
functional derivatives:
\begin{align}
  \fv_\rho^\rmsup(\rv,t;[\rho,G,\Jv,J_G]) &= 
  T \frac{\delta \Sigma_t^\rmexc[\rho,G,\Jv,J_G]}{\delta \Jv(\rv,t)},
  \label{EQfsupRho}\\
  f_G^\rmsup(r,t;[\rho,G,\Jv,J_G]) &=
  T \frac{\delta \Sigma_t^\rmexc[\rho,G,\Jv,J_G]}{\delta J_G(r,t)}.
  \label{EQfsupG}
\end{align}

It is useful to combine the two adiabatic force fields
\eqref{EQfadRho} and \eqref{EQfadG} each with their respective
superadiabatic partner~\eqref{EQfsupRho} and \eqref{EQfsupG} to obtain
intrinsic (entropic excess) force functionals for position motion,
$\fv_\rho^\rmint(\rv,t;[\rho,G,\Jv,J_G])$, and for distance motion,
$f_G^\rmint(r,t;[\rho,G,\Jv,J_G])$. Explicitly, these are given as the
following sums:
\begin{align}
  \fv_\rho^\rmint(\rv,t;[\rho,G,\Jv,J_G]) &=
  \fv_\rho^\rmad(\rv,t;[\rho,G])\notag\\&\quad
  + \fv_\rho^\rmsup(\rv,t;[\rho,G,\Jv,J_G]),
  \label{EQfintRho}\\
  f_G^\rmint(r,t;[\rho,G,\Jv,J_G]) &= 
  f_G^\rmad(r,t;[\rho,G]) \notag\\&\quad
  + f_G^\rmsup(r,t;[\rho,G,\Jv,J_G]).
  \label{EQfintG}
\end{align}

We can now make the Euler-Lagrange equations \eqref{EQpowerEL1} and
\eqref{EQpowerEL2} more explicit upon inserting the free power
decompositions \eqref{EQRtDefinition} and \eqref{EQWtDefinition}
together with the force results from functional differentiation, as
described below.  This yields the following pair of formally exact
coupled functional equations of motion:
\begin{align}
  \gamma \vel_\rho^\rmid(\rv,t;[\rho,G,J_G]) &= 
  -k_BT\nabla \ln\rho(\rv,t) 
  + \fext(\rv,t)
  \notag\\&\qquad
  + \fv_\rho^\rmint(\rv,t;[\rho,G,\Jv,J_G]),
  \label{EQofMotion1}  \\
  \gamma J_G(r,t)/G(r,t)
  &=  \gamma v_G^\rmid(r,t;[\rho,G,\Jv])  -\pairpotential'(r,t)
  \notag\\&\qquad
  + f_G^\rmint(r,t;[\rho,G,\Jv,J_G]).
  \label{EQofMotion2}
\end{align}
We have used the potential forces
\eqref{EQexternalForceFromDifferentiation} and
\eqref{EQinterparticleForceFromDifferentiation}, the diffusive force
\eqref{EQdiffusiveForceFromDifferentiation}, the friction terms
\eqref{EQPtidByPosition} and \eqref{EQPtidByDistance}, as well as the
interparticle functionals \eqref{EQfintRho} and~\eqref{EQfintG}.
We recall $v_G(r,t)$ as the ratio \eqref{EQvGDefinition} and the
ideal velocity functionals $v_G^\rmid(r,t;[\rho,G,\Jv])$ and
$\vel_\rho^\rmid(\rv,t;[\rho,G,J_G])$ as the respective explicit
functionals \eqref{EQvidGDefinition} and \eqref{EQvelIdealDefinition}.

Together with the continuity equations \eqref{EQcontinuity1} and
\eqref{EQcontinuity2}, the equations of motion \eqref{EQofMotion1} and
\eqref{EQofMotion2} determine formally the dynamics of the system.
The many-body problem is encapsulated in the adiabatic entropy
metadensity functional $S_t^\rmexc[\rho,G]$ and in the entropic
superpower functional $\Sigma_t^\rmexc[\rho,G,\Jv,J_G]$. The
dependence on the external force field $\fext(\rv,t)$ occurs solely
and explicitly in the position equation of motion
\eqref{EQofMotion1}. Similarly, and arguably more strikingly, the
dependence on the pair potential $\pairpotential(r)$ is merely via the
explict occurrence of the pair force $-\phi'(r)$ in the distance
equation of motion \eqref{EQofMotion2}.

As a consistency check we address the free diffusion of the ideal gas,
where we have $\fext=\fv_\rho^\rmint=0$ in the position equation of
motion \eqref{EQofMotion1} and $\pairpotential'=f_G^\rmint=0$ in the
distance equation of motion \eqref{EQofMotion2}.  The ideal distance
distribution is $G_\rmid(r,t)=\int d\rv
d\rv'\rho(\rv,t)\rho(\rv',t)\delta(r-|\rv-\rv'|)/2$ and it is
straightforward to show that $\dot G_\rmid(r,t) = -\partial_r
J_G^\rmid(r,t)$ which is consistent with $J_G(r,t) = J_G^\rmid(r,t)$,
as follows from the equation of motion
\eqref{EQofMotion1}. Furthermore $\vel(\rv,t) =
\vel_\rho^\rmid(\rv,t)$, such that from the equation of
motion~\eqref{EQofMotion1} one obtains $\gamma
\vel(\rv,t)=-k_BT\nabla\ln\rho(\rv,t)$, which is exact for the ideal
gas. The present work offers a systematic means to go further and
encapsulate all many-body correlation effects that drive the full
nonequilibrium dynamics in an efficient functional framework.

As an outlook on future work, it would be highly interesting to carry
out analytical work to develop superpower approximations, to construct
machine learning schemes based on the relevant functional maps, and to
explore the interconnections with statistical mechanical gauge
invariance \cite{mueller2024dynamic}, with the fluctuation theorems of
stochastic thermodynamics \cite{seifert2012}, with recent
trajectory-based work~\cite{deguenther2024, meyberg2024}, and with
descriptions of active interaction switching theory
\cite{monchojorda2020}.
Furthermore, whether additional insights could be gained for recently
identified collective effects in Brownian dynamics, such as the
emergence of solitons \cite{antonov2022,ceredalopz2023} and fractional
Shapiro steps in colloidal transport \cite{stikuts2025, mishra2025}
constitute highly stimulating questions.

\bigskip
{\bf Acknowledgments.}  This work is supported by the DFG (Deutsche
Forschungsgemeinschaft) under project no.~523317330.

\bibliographystyle{prsty}
\bibliography{noe}

\begin{thebibliography}{10}

\bibitem{schmidt2022rmp}
M. Schmidt, {Power functional theory for many-body dynamics},
  \href{https://doi.org/10.1103/RevModPhys.94.015007} {Rev. Mod. Phys. {\bf
  94}, 015007 (2022).}

\bibitem{brader2015dtpl}
J. M. Brader and M. Schmidt, {Power functional theory for the dynamic test
  particle limit,} \href{https://doi.org/10.1088/0953-8984/27/19/194106} {J.
  Phys.: Condens. Matter {\bf 27}, 194106 (2015).}

\bibitem{treffenstaedt2021dtpl}
L. L. Treffenst\"adt and M. Schmidt, {Universality in driven and equilibrium
  hard sphere liquid dynamics,}
  \href{https://doi.org/10.1103/PhysRevLett.126.058002} {Phys. Rev. Lett. {\bf
  126}, 058002 (2021).}

\bibitem{treffenstaedt2022dtpl}
L. L. Treffenst\"adt, T. Schindler, M. Schmidt, {Dynamic decay and
  superadiabatic forces in the van Hove dynamics of bulk hard sphere fluids},
  \href{https://doi.org/10.21468/SciPostPhys.12.4.133} {SciPost Phys. {\bf 12},
  133 (2022).}

\bibitem{brader2013noz}
J. M. Brader and M. Schmidt, Dynamic correlations in Brownian many-body
  systems, \href{http://dx.doi.org/10.1063/1.4861041} {J. Chem. Phys. {\bf
  140}, 034104 (2014).}

\bibitem{brader2014noz}
J. M. Brader and M. Schmidt, Nonequilibrium Ornstein-Zernike relation for
  Brownian many-body dynamics, \href{http://dx.doi.org/10.1063/1.4820399} {J.
  Chem. Phys. {\bf 139}, 104108 (2013).}

\bibitem{schmidt2013pft}
M. Schmidt and J. M. Brader, {Power functional theory for Brownian dynamics},
  \href{https://doi.org/10.1063/1.4807586} {J. Chem. Phys. {\bf 138}, 214101
  (2013).}

\bibitem{schmidt2018md}
M. Schmidt, {Power functional theory for Newtonian many-body dynamics,}
  \href{https://doi.org/10.1063/1.5008608} {J. Chem. Phys. {\bf 148}, 044502
  (2018).}

\bibitem{schmidt2015qpft}
M. Schmidt, {Quantum power functional theory for many-body dynamics,}
  \href{https://doi.org/10.1063/1.4934881} {J. Chem. Phys. {\bf 143}, 174108
  (2015).}

\bibitem{hansen2013}
J.~P. Hansen and I.~R. McDonald, {\it Theory of Simple Liquids}, 4th ed.\
  (Academic Press, London, 2013).

\bibitem{evans1979}
R. Evans, {The nature of the liquid-vapour interface and other topics in the
  statistical mechanics of non-uniform, classical fluids},
  \href{https://doi.org/10.1080/00018737900101365} {Adv. Phys. {\bf 28}, 143
  (1979).}

\bibitem{evans1992}
R. Evans, {Density functionals in the theory of nonuniform fluids},
  \href{https://books.google.com/books?id=-fNr2a4v3bYC&pg=PA85} {Chap.~3 in
  {\it Fundamentals of Inhomogeneous Fluids}, edited by D. Henderson (Dekker,
  New York, 1992).}

\bibitem{evans2016}
R. Evans, M. Oettel, R. Roth, and G. Kahl, {New developments in classical
  density functional theory},
  \href{https://doi.org/10.1088/0953-8984/28/24/240401} {J. Phys.: Condens.
  Matter {\bf 28}, 240401 (2016).}

\bibitem{fortini2014prl}
A. Fortini, D. de las Heras, J.~M. Brader, and M.~Schmidt, {Superadiabatic
  forces in Brownian many-body dynamics,}
  \href{https://doi.org/10.1103/PhysRevLett.113.167801} {Phys. Rev. Lett. {\bf
  113}, 167801 (2014).}

\bibitem{bernreuther2016gcm}
E. Bernreuther and M. Schmidt, Superadiabatic forces in the dynamics of the
  one-dimensional Gaussian core model,
  \href{http://dx.doi.org/10.1103/PhysRevE.94.022105} {Phys. Rev. E {\bf 94},
  022105 (2016).}

\bibitem{delasheras2018velocityGradient}
D. de las Heras and M. Schmidt, {Velocity gradient power functional for
  Brownian dynamics}, \href{https://doi.org/10.1103/PhysRevLett.120.028001}
  {Phys. Rev. Lett. {\bf 120}, 028001 (2018).}

\bibitem{stuhlmueller2018prl}
N. C. X. Stuhlm\"uller, T. Eckert, D. de las Heras, and M. Schmidt, {Structural
  nonequilibrium forces in driven colloidal systems,}
  \href{https://doi.org/10.1103/PhysRevLett.121.098002} {Phys. Rev. Lett. {\bf
  121}, 098002 (2018).}

\bibitem{delasheras2020fourForces}
D. de las Heras and M. Schmidt, {Flow and structure in nonequilibrium Brownian
  many-body systems}, \href{https://doi.org/10.1103/PhysRevLett.125.018001}
  {Phys. Rev. Lett. {\bf 125}, 018001 (2020).}

\bibitem{treffenstaedt2020shear}
L. L. Treffenst\"adt and M. Schmidt, {Memory-induced motion reversal in
  Brownian liquids,} \href{https://doi.org/10.1039/C9SM02005E} {Soft Matter
  {\bf 16}, 1518 (2020).}

\bibitem{jahreis2019shear}
N. Jahreis and M. Schmidt, {Shear-induced deconfinement of hard disks,}
  \href{https://doi.org/10.1007/s00396-020-04644-1} {Col. Pol. Sci. {\bf 298},
  895 (2020).}

\bibitem{geigenfeind2020laning}
T. Geigenfeind, D. de las Heras and M. Schmidt, Superadiabatic demixing in
  nonequilibrium colloids, \href{https://doi.org/10.1038/s42005-020-0287-5}
  {Commun. Phys. {\bf 3}, 23 (2020).}

\bibitem{koeglmayr2024scaling}
J. K\"oglmayr, F. Samm\"uller, and M. Schmidt, Nonequilibrium scaling of drag
  forces in counterdriven fluid mixtures,
  \href{https://doi.org/10.48550/arXiv.2605.31479} {arXiv:2605.31479.}

\bibitem{krinninger2016prl}
P. Krinninger, M. Schmidt, and J. M. Brader, {Nonequilibrium phase behaviour
  from minimization of free power dissipation},
  \href{http://dx.doi.org/10.1103/PhysRevLett.117.208003} {Phys. Rev. Lett.
  {\bf 117}, 208003 (2016).}

\bibitem{krinninger2019jcp}
P. Krinninger and M. Schmidt, {Power functional theory for active Brownian
  particles: general formulation and power sum rules,}
  \href{https://doi.org/10.1063/1.5061764} {J. Chem. Phys. {\bf 150}, 074112
  (2019).}

\bibitem{hermann2024activeFreezing}
S. Hermann and M. Schmidt, {Active crystallization from power functional
  theory,} \href{https://doi.org/10.1103/PhysRevE.109.L022601} {Phys. Rev. E
  (Letter) {\bf 109}, L022601 (2024).}

\bibitem{hermann2019prl}
S. Hermann, D. de las Heras, and M. Schmidt, {Non-negative interfacial tension
  in phase-separated active Brownian particles,}
  \href{https://doi.org/10.1103/PhysRevLett.123.268002} {Phys. Rev. Lett. {\bf
  123}, 268002 (2019).}

\bibitem{hermann2021molPhys}
S. Hermann, D. de las Heras, and M. Schmidt, {Phase separation of active
  Brownian particles in two dimensions: Anything for a quiet life,}
  \href{https://doi.org/10.1080/00268976.2021.1902585} {Mol. Phys. e1902585
  (2021).}

\bibitem{bruetting2019viscosity}
M. Br\"utting, T. Trepl, D. de las Heras, and M. Schmidt, {Superadiabatic
  forces via the acceleration gradient in quantum many-body dynamics,}
  \href{https://doi.org/10.3390/molecules24203660} {Molecules {\bf 24}, 3660
  (2019).}

\bibitem{renner2022prl}
J. Renner, M. Schmidt, and D. de las Heras, {Shear and bulk acceleration
  viscosities in simple fluids,}
  \href{https://doi.org/10.1103/PhysRevLett.128.094502} {Phys. Rev. Lett. {\bf
  128}, 094502 (2022).}

\bibitem{mermin1965}
N. D. Mermin, Thermal properties of the inhomogeneous electron gas,
  \href{https://doi.org/10.1103/PhysRev.137.A1441} {Phys. Rev. {\bf 137}, A1441
  (1965).}

\bibitem{delasheras2019customFlow}
D. de las Heras, J. Renner, and M. Schmidt, {Custom flow in overdamped Brownian
  dynamics,} \href{https://doi.org/10.1103/PhysRevE.99.023306} {Phys. Rev. E
  {\bf 99}, 023306 (2019).}

\bibitem{renner2021customFlowMD}
J. Renner, M. Schmidt, and D. de las Heras, {Custom flow in molecular
  dynamics,} \href{https://doi.org/10.1103/PhysRevResearch.3.013281} {Phys.
  Rev. Research {\bf 3}, 013281 (2021).}

\bibitem{delasheras2023perspective}
D. de las Heras, T. Zimmermann, F. Samm\"uller, S. Hermann, and M. Schmidt,
  Perspective: How to overcome dynamical density functional theory,
  \href{https://doi.org/10.1088/1361-648X/accb33} {J. Phys.: Condens. Matter
  {\bf 35}, 271501 (2023); (Invited Perspective)}.

\bibitem{zimmermann2024ml}
T. Zimmermann, F. Samm\"uller, S. Hermann, M. Schmidt, and D. de las Heras,
  Neural force functional for non-equilibrium many-body colloidal systems,
  \href{https://doi.org/10.1088/2632-2153/ad7191} {Mach. Learn.: Sci. Technol.
  {\bf 5}, 035062 (2024).}

\bibitem{fleischmann2026}
P. Fleischmann, M. Schmidt, and F. Samm\"uller, Power functional learning the
  physics of active bulk fluids (to be published).

\bibitem{sammueller2023neural}
F. Samm\"uller, S. Hermann, D. de las Heras, and M. Schmidt, {Neural functional
  theory for inhomogeneous fluids: Fundamentals and applications},
  \href{https://doi.org/10.1073/pnas.2312484120} {Proc. Natl. Acad. Sci. {\bf
  120}, e2312484120 (2023).}

\bibitem{dijkman2024ml}
J. Dijkman, M. Dijkstra, R. van Roij, M. Welling, J.-W. van de Meent, and B.
  Ensing, Learning neural free-energy functionals with pair-correlation
  matching, \href{https://doi.org/10.1103/PhysRevLett.134.056103} {Phys. Rev.
  Lett. {\bf 134}, 056103 (2025).}

\bibitem{sammueller2024attraction}
F. Samm\"uller, M. Schmidt, and R. Evans, Neural density functional theory of
  liquid-gas phase coexistence,
  \href{https://doi.org/10.1103/PhysRevX.15.011013} {Phys. Rev. X {\bf 15},
  011013 (2025)}; \href{https://doi.org/10.1103/Physics.18.17} {Featured in
  Physics {\bf 18}, 17 (2025)}.

\bibitem{bui2024neuralrpm}
A. T. Bui and S. J. Cox, Learning classical density functionals for ionic
  fluids, \href{https://doi.org/10.1103/PhysRevLett.134.148001} {Phys. Rev.
  Lett. {\bf 134}, 148001 (2025).}

\bibitem{bui2025dielectrocapillarity}
A. T. Bui and S. J. Cox, Dielectrocapillarity for exquisite control of fluids,
  \href{https://doi.org/10.1038/s41467-026-69482-1} {Nat. Commmun. {\bf 17},
  2661 (2026).}

\bibitem{bui2026abinitio}
A. T. Bui and S. J. Cox, A unified machine learning framework for ab initio
  multiscale modeling of liquids,
  \href{https://doi.org/10.48550/arXiv.2603.20493} {arXiv:2603.20493.}

\bibitem{sammueller2024hyperDFT}
F. Samm\"uller, S. Robitschko, S. Hermann, and M. Schmidt, Hyperdensity
  functional theory of soft matter,
  \href{https://doi.org/10.1103/PhysRevLett.133.098201} {Phys. Rev. Lett. {\bf
  133}, 098201 (2024); PRL Editors' Suggestion.}

\bibitem{kampa2024meta}
S. M. Kampa, F. Samm\"uller, M. Schmidt, and R. Evans, Metadensity functional
  theory for classical fluids: Extracting the pair potential,
  \href{https://doi.org/10.1103/PhysRevLett.134.107301} {Phys. Rev. Lett. {\bf
  134}, 107301 (2025);} PRL Editors' Suggestion.

\bibitem{kampa2026pairmatching}
S. M. Kampa, F. Samm\"uller, and M. Schmidt, Metadensity functional learning
  for classical fluids: Regularizing with pair correlations,
  \href{https://doi.org/10.1021/acs.jpcb.6c01662} {J. Phys. Chem. B {\bf 130},
  6231 (2026).} (Special issue: {\it Classical density functional theory in
  physical chemistry}).

\bibitem{kampa2026spherical}
S. M. Kampa, M. Schmidt, and F. Samm\"uller, Spherical metadensity functional
  learning for inhomogeneous classical fluids,
  \href{https://doi.org/10.48550/arXiv.2606.14370} {arXiv:2606.14370.}

\bibitem{marconi1999}
U. M. B. Marconi and P. Tarazona, {Dynamic density functional theory of
  fluids}, \href{https://doi.org/10.1063/1.478705} {J. Chem. Phys. {\bf 110},
  8032 (1999).}

\bibitem{marconi2007}
U. M. B. Marconi and S. Melchionna, {Phase-space approach to dynamical density
  functional theory}, \href{https://doi.org/10.1063/1.2724823} {J. Chem. Phys.
  {\bf 126}, 184109 (2007).}

\bibitem{schilling2022}
T. Schilling, {Coarse-grained modelling out of equilibrium,}
  \href{https://doi.org/10.1016/j.physrep.2022.04.006} {Phys. Rep. {\bf 972}, 1
  (2022).}

\bibitem{noether1918}
E. Noether, {Invariante Variationsprobleme,}
  \href{https://gdz.sub.uni-goettingen.de/download/pdf/PPN252457811_1918/LOG_0022.pdf}
  {Nachr. d. K\"onig. Gesellsch. d. Wiss. zu G\"ottingen, Math.-Phys. Klasse,
  {\bf 235}, 183 (1918).} English translation by M. A. Tavel: Invariant
  variation problems. \href{https://doi.org/10.1080/00411457108231446} {Transp.
  Theo. Stat. Phys. {\bf 1}, 186 (1971)}; for a version in modern typesetting
  see: Frank Y. Wang,
  \href{http://arxiv.org/abs/physics/0503066v3}{arXiv:physics/0503066v3}
  (2018).

\bibitem{hermann2021noether}
S. Hermann and M. Schmidt, {Noether's theorem in statistical mechanics},
  \href{https://doi.org/10.1038/s42005-021-00669-2} {Commun. Phys. {\bf 4}, 176
  (2021).}

\bibitem{mueller2024gauge}
J. M\"uller, S. Hermann, F. Samm\"uller, and M. Schmidt, Gauge invariance of
  equilibrium statistical mechanics,
  \href{https://doi.org/10.1103/PhysRevLett.133.217101} {Phys. Rev. Lett. {\bf
  133}, 217101 (2024)}; Editors' Suggestion; PRL's
  \href{https://promo.aps.org/PRL2024} {Collection of the Year} 2024; Featured
  in \href{https://doi.org/10.1103/Physics.17.163} {Physics {\bf 17}, 163
  (2024)} by B. Rotenberg.

\bibitem{mueller2024whygauge}
J. M\"uller, F. Samm\"uller, and M. Schmidt, Why gauge invariance applies to
  statistical mechanics, \href{https://doi.org/10.1088/1751-8121/adbfe6} {J.
  Phys. A: Math. Theor. {\bf 58}, 125003 (2025).}

\bibitem{mueller2024dynamic}
J. M\"uller, F. Samm\"uller, and M. Schmidt, Dynamical gauge invariance of
  statistical mechanics, \href{https://doi.org/10.48550/arXiv.2504.17599}
  {arXiv:2504.17599.}

\bibitem{schmidt2026entropyShort}
M. Schmidt, Entropy density functional theory for inhomogeneous fluids,
  \href{https://doi.org/10.48550/arXiv.2606.28240} {arXiv:2606.28240.}

\bibitem{schmidt2026entropyLong}
M. Schmidt, Entropy density functional universality: Correlation, response, and
  entropic Ornstein-Zernike structure,
  \href{https://doi.org/10.48550/arXiv.2607.03032} {arXiv:2607.03032.}

\bibitem{zwanzig2001}
R. Zwanzig, {\it Nonequilibrium Statistical Mechanics} (Oxford University
  Press, Oxford, 2001).

\bibitem{dhont1996book}
J. K. G. Dhont, An Introduction to the Dynamics of Colloids (Elsevier,
  Amsterdam, 1996).

\bibitem{seifert2012}
U. Seifert, {Stochastic thermodynamics, fluctuation theorems and molecular
  machines,} \href{https://doi.org/10.1088/0034-4885/75/12/126001} {Rep. Prog.
  Phys. {\bf 75}, 126001 (2012).}

\bibitem{deguenther2024}
J. Deg\"unther, J. van der Meer, and U. Seifert, General theory for localizing
  the where and when of entropy production meets single-molecule experiments
  \href{https://doi.org/10.1073/pnas.2405371121 } {Proc. Natl. Acad. Sci. {\bf
  120}, e2405371121 (2024).}

\bibitem{meyberg2024}
E. Meyberg, J. Deg\"unther, and U. Seifert, Entropy production from
  waiting-time distributions for overdamped Langevin dynamics,
  \href{https://doi.org/10.1088/1751-8121/ad508a} {J. Phys. A: Math. Theor.
  {\bf 57} 25LT01 (2024).}

\bibitem{monchojorda2020}
A. Moncho-Jord\'a and J. Dzubiella, {Controlling the microstructure and phase
  behavior of confined soft colloids by active interaction switching,}
  \href{https://doi.org/10.1103/PhysRevLett.125.078001} {Phys. Rev. Lett. {\bf
  125}, 078001 (2020).}

\bibitem{antonov2022}
A. P. Antonov, A. Ryabov, and P. Maass, {Solitons in overdamped Brownian
  dynamics,} \href{https://doi.org/10.1103/PhysRevLett.129.080601} {Phys. Rev.
  Lett. {\bf 129}, 080601 (2022).}

\bibitem{ceredalopz2023}
E. Cereceda-L\'opez, A. P. Antonov, A. Ryabov, P. Maass, and P. Tierno,
  Overcrowding induces fast colloidal solitons in a slowly rotating potential
  landscape \href{https://doi.org/10.1038/s41467-023-41989-x} {Nat. Commun.
  {\bf 14}, 6448 (2023).}

\bibitem{stikuts2025}
A. P. Stikuts, S. Mishra, A. Ryabov, P. Maass, and P. Tierno, Engineering
  tunable fractional Shapiro steps in colloidal transport,
  \href{https://doi.org/10.1038/s41467-025-58217-3} {Nat. Commun. {\bf 16},
  2966 (2025).}

\bibitem{mishra2025}
S. Mishra, A. Ryabov, and P. Maass, Phase locking and fractional Shapiro steps
  in collective dynamics of microparticles,
  \href{https://doi.org/10.1103/PhysRevLett.134.107102} {Phys. Rev. Lett. {\bf
  134}, 107102 (2025).}

\end{thebibliography}

\end{document}